\documentclass[twocolumn,tighten,times]{aastex62}

\usepackage{natbib}
\usepackage{MnSymbol}
\usepackage{multirow}
\usepackage{gensymb}
\usepackage{amsmath}
\usepackage{soul}

\newcommand{\cena}{NGC\,5128}
\newcommand{\mstar}{${\cal M}_\star$}
\newcommand{\reff}{$r_{\rm eff}$}

\received{\today}
\submitjournal{ApJL}

\shorttitle{Dwarf Galaxies in \cena's Western Halo}
\shortauthors{Taylor et al.}

\begin{document}

\title{A Collection of New Dwarf Galaxies in \cena's Western Halo}

\correspondingauthor{Matthew A.\ Taylor}
\email{mtaylor@gemini.edu}


\author[0000-0003-3009-4928]{Matthew A. Taylor}
\affil{Gemini Observatory, Northern Operations Center, 670 North A'ohoku Place, Hilo, HI 96720-1906, USA}
\author[0000-0001-8654-0101]{Paul Eigenthaler}
\affiliation{Instituto de Astrof\'isica, Pontificia Universidad Cat\'olica de Chile, Av.~Vicu\~na Mackenna 4860, 7820436 Macul, Santiago, Chile}
\author[0000-0003-0350-7061]{Thomas H. Puzia}
\affiliation{Instituto de Astrof\'isica, Pontificia Universidad Cat\'olica de Chile, Av.~Vicu\~na Mackenna 4860, 7820436 Macul, Santiago, Chile}
\author[0000-0003-1743-0456]{Roberto P. Mu\~noz}
\affiliation{Instituto de Astrof\'isica, Pontificia Universidad Cat\'olica de Chile, Av.~Vicu\~na Mackenna 4860, 7820436 Macul, Santiago, Chile}
\author[0000-0002-3004-4317]{Karen X. Ribbeck}
\affiliation{Instituto de Astrof\'isica, Pontificia Universidad Cat\'olica de Chile, Av.~Vicu\~na Mackenna 4860, 7820436 Macul, Santiago, Chile}
\author[0000-0003-1632-2541]{Hong-Xin Zhang}
\affiliation{CAS Key Laboratory for Research in Galaxies and Cosmology, Department of Astronomy, University of Science and Technology of China, Hefei 230026, China}
\author[0000-0001-7966-7606]{Yasna Ordenes-Brice\~no}
\affiliation{Instituto de Astrof\'isica, Pontificia Universidad Cat\'olica de Chile, Av.~Vicu\~na Mackenna 4860, 7820436 Macul, Santiago, Chile}
\author{Mia Sauda Bovill}
\affiliation{Texas Christian University, Dept.\ of Physics and Astronomy, 2800 S.\ University Dr., Fort Worth, TX 76129, USA}

\begin{abstract}
We report the photometric properties of 16 dwarf galaxies, 15 of which are newly identified, in the Western halo of the nearby giant elliptical galaxy \cena.~All candidates are found at projected distances $\sim\!100\!-\!225$\,kpc from their giant host, with luminosities $-10.82\!\leq\!M_V\!/{\rm mag}\!\leq\!-7.42$ and effective radii $4\arcsec\!\la\!r_{\rm eff}\!\la\!17\arcsec$ (or $75\!\la\!r_{\rm eff}\!/{\rm pc}\!\la\!300$ at the distance of \cena).~We compare to other low-mass dwarf galaxies in the local universe and find them to populate the faint/compact extension of the size-luminosity relation that was previously not well-sampled by dwarf galaxies in the Centaurus\,A system, with optical colors similar to compact stellar systems like globular clusters and ultra-compact dwarf galaxies despite having much more diffuse morphologies. From optical $u'g'r'i'z'$ photometry, stellar masses are estimated to be $5.17\!\leq\!\log$\mstar$\!/\!M_\odot\!\leq\!6.48$, with colors that show them to fall redward of the dwarf galaxy mass-metallicity relation. These colors suggest star formation histories that require some mechanism that would give rise to extra metal enrichment such as primordial formation within the halos of their giant galaxy hosts, non-primordial star formation from previously enriched gas, or extended periods of star formation leading to self-enrichment.~We also report the existence of at least two sub-groups of dwarf candidates, each subtending $\la\!15\arcmin$ on the sky, corresponding to projected physical separations of $10\!-\!20$\,kpc.~True physical associations of these groups, combined with their potentially extended star formation histories, would imply that they may represent dwarf galaxy groups in the early stage of interaction upon infall into a giant elliptical galaxy halo in the very nearby universe.
\end{abstract}

\keywords{galaxies: individual (NGC\,5128) --- galaxies: dwarf --- galaxies: elliptical and lenticular, cD}

\section{Introduction} \label{sec:intro}
A renaissance is underway regarding the discovery rate of low-luminosity dwarf galaxies throughout the local universe at distances $\la50$\,Mpc.~The known population of dwarf galaxies in the Local Group (LG) has risen in the last decade to include dozens \citep[e.g.][]{bel10, mcc12, bec15, mun18a, mun18b}, while rich dwarf galaxy systems beyond the LG have been discovered in myriad environments including galaxy clusters like Virgo and Fornax \citep[][]{san16, mun15, eig18, ord18a} and smaller complexes including M81, M101, NGC\,2784, M96, and others \citep[e.g.][]{chi09, mer14, jav16, ben17, hen17, mul17b, sme17, par17, mul18a}.~It becomes increasingly apparent that such dwarf galaxy systems exhibit seemingly structured distributions around their host gravitational potentials, with surprisingly thin, planar distributions found around the LG \citep[][]{iba13, paw12}, and evidence for sub-$100$\,kpc-scale clustering for dwarfs orbiting within the half-virial sphere of the Fornax galaxy cluster \citep[][]{mun15, ord18a}.

The nearby giant elliptical galaxy \cena\ has proven to be no exception to the above findings.~Recent years have seen the discovery of several new dwarf galaxy satellites, including the existence of a co-rotating plane of satellites surrounding \cena\ itself \citep[][]{mul18b}, numerous dwarf galaxies amongst tidal debris in \cena's Northeastern halo \citep[][]{crn14, crn16}, which is all part of an apparent ``bridge'' of satellites spread Northward from \cena, towards the associated giant galaxy M83 \citep[][]{mul15, mul16, mul17a}.~Meanwhile, the globular cluster (GC) system of \cena\ has been found to extend in the 1000s to at least $200$\,kpc---and very likely beyond---which is dominated by blue, presumably metal-poor, GCs \citep[][]{tay17}.~The existence of such a rich population of blue GCs may suggest the existence of a former and/or present reservoir of low-mass dwarf galaxy GC hosts.~Given the above, there is a natural motivation to search for yet more dwarf galaxies in \cena's halo, particularly in the Northern regions where higher numbers of dwarfs might be expected to accompany the \cena-M83 bridge.

We focus in this work on a region $\sim\!100-225$\,kpc into the North/Western halo, where little deep imaging data has been obtained to date.~In what follows we report the properties of 16 dwarf galaxy candidates (15 new).~We determine multi-band luminosities and structural parameters, which are used to broadly compare to other low-mass systems throughout the local universe and to predictions of simple stellar population (SSP) evolutionary model tracks.~Throughout this work, we adopt a \cena\ distance modulus of $27.88$ mag \citep[$3.8$\,Mpc][]{har10}, corresponding to a spatial scale of $19$\,pc arcsec$^{-1}$.

\begin{figure*}
\includegraphics[width=\linewidth]{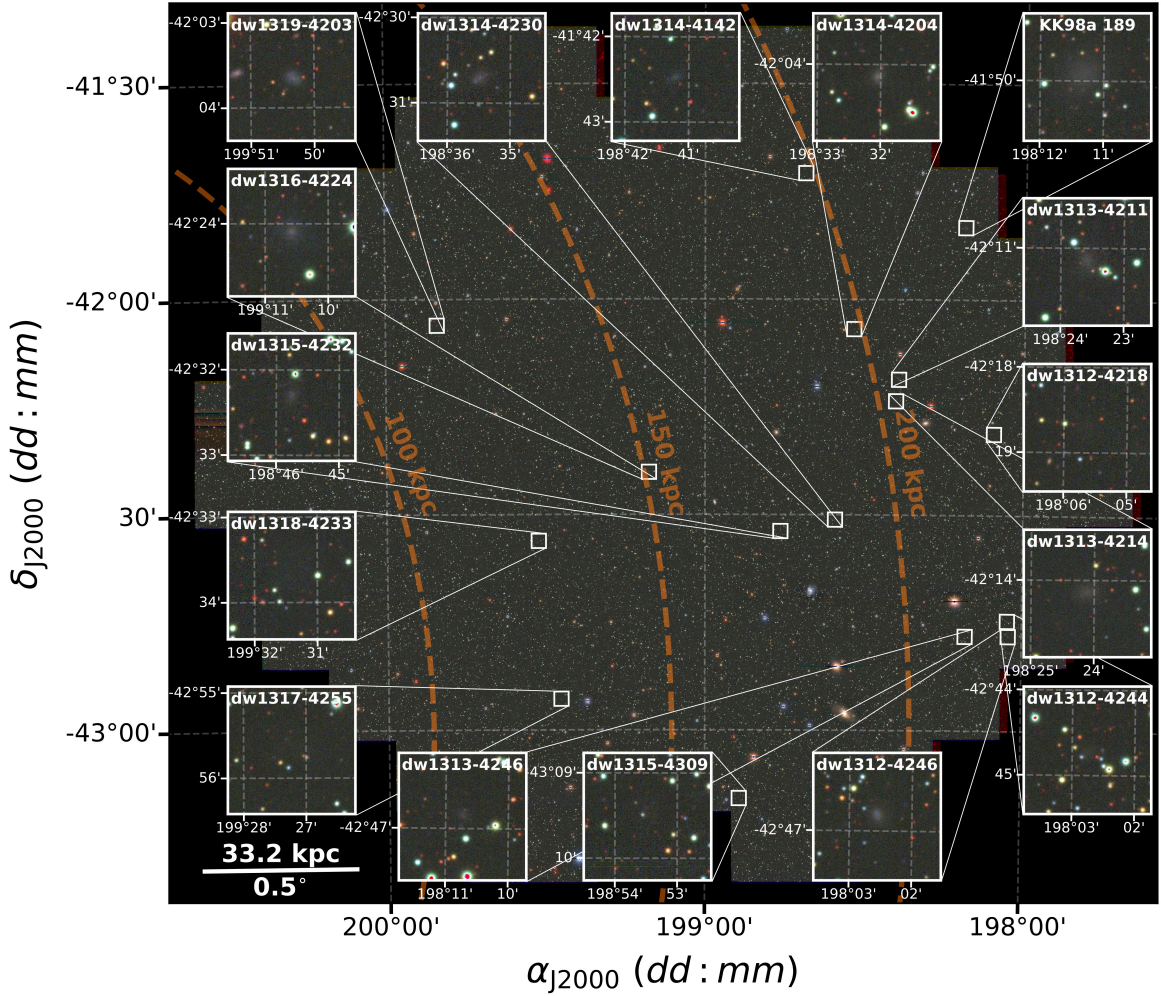}
\caption{Locations of the new dwarf galaxy candidates on a $u'g'z'$-based RGB color image.~White-bordered images show $1.5\arcmin\!\times\!1.5\arcmin$ cutouts centered on the dwarfs, corresponding to $\sim\!1.7\!\times\!1.7$\,kpc$^2$ at the distance of \cena.~The scale of the background frame is indicated in the lower left corner.~Dashed orange curves show distances corresponding to 100, 150, and 200\,kpc from \cena.~North is up, East is to the left.} 
\label{fig:dw_images}
\end{figure*} 

\section{Observations and Image Processing} \label{sec:data}
Observations in the optical $u'g'r'i'z'$ filters were obtained as part of the {\it Survey of Centaurus A's Baryonic Structures} campaign \citep[][]{tay17} that uses the {\it Dark Energy Camera} (DECam) mounted on the 4m-Blanco telescope at the Cerro Tololo Interamerican Observatory (CTIO) to image $\sim\!72$\,deg$^2$\ of the sky around \cena.~Specifically we utilize, for each filter, a set of DECam images centered on $(\alpha,\delta)=(13\!:\!16\!:\!30.83, -42\!:\!21\!:\!41.06)$.~Images were acquired over 4-5 April 2014 with total integration times of $1200$, $100$, $60$, $100$, and $200$\,s targeted to reach S/N~$\!\approx\!5$ point-source depths of $24.1$, $22.7$, $22.5$, $22.1$, and $21.7$ in the $u'$, $g'$, $r'$, $i'$, and $z'$ filters, respectively. Image preprocessing was carried out by the DECam community pipeline \citep[CP;][]{val14} to remove instrumental signatures (bias subtraction, flat-fielding, cross-talk correction, fringing, etc.).~From the CP calibrated frames, we use the {\sc Astromatic}\footnote{\url{http://www.astromatic.net/}} software suite \citep[Source Extractor, hereafter SE; SCAMP; SWARP; PSFEx;][]{ber96,ber02,ber06} to register frames to a common coordinate system and account for pixel scale distortions across the DECam field of view using the 2MASS astrometric reference star catalogue and calibrate our photometry to the SDSS system using frames of the LSE\_44 SDSS southern standard star field taken at varying airmass during the observing nights.

\section{Analysis}
\label{sec:analysis}
\subsection{Dwarf Detection and Photometric Analysis}
\label{sec:detect}
We detect dwarf candidates using a full-color RGB image constructed from our $u'g'z'$ imaging with custom {\sc Python} image processing scripts.~This filter combination samples the full optical spectral energy distribution (SED) and is sensitive to old, metal-poor stellar populations expected of primordial dwarf galaxies, while the inclusion of the $u'$-band also serves to sample flux from younger stellar components arising from more recent star-formation.~Figure\,\ref{fig:dw_images} shows the RGB image with orange dashed-curves indicating \cena-centric iso-radial contours.~We visually searched this image for faint, extended sources displaying smooth morphologies with shallow surface brightness ($\mu$) profiles typical of low-luminosity dwarfs.~Using this method, we unambiguously find 16 dwarf galaxy candidates, of which a single previously known dwarf is recovered \citep[KK98a\,189;][]{kar98} and the remaining 15 are reported here for the first time.~The set of zoom-in images in Figure\,\ref{fig:dw_images} show the dwarf candidates, and measure $2\arcmin$ on a side, corresponding to physical $\sim2.2\times2.2\,{\rm kpc}^2$ regions at the distance of \cena.~We point out here, that by employing a ``by eye'' detection technique, we are unable to make a robust estimate of the completeness limit to our imaging in a straightforward manner and thus cannot make broad statements on the true total population of dwarfs in this region.~We instead defer such an analysis to a future work detailing the overall properties of dwarfs detected throughout our survey region, noting that a lack of completeness estimate does not affect our conclusions.

\begin{figure*}
\includegraphics[width=0.95\linewidth]{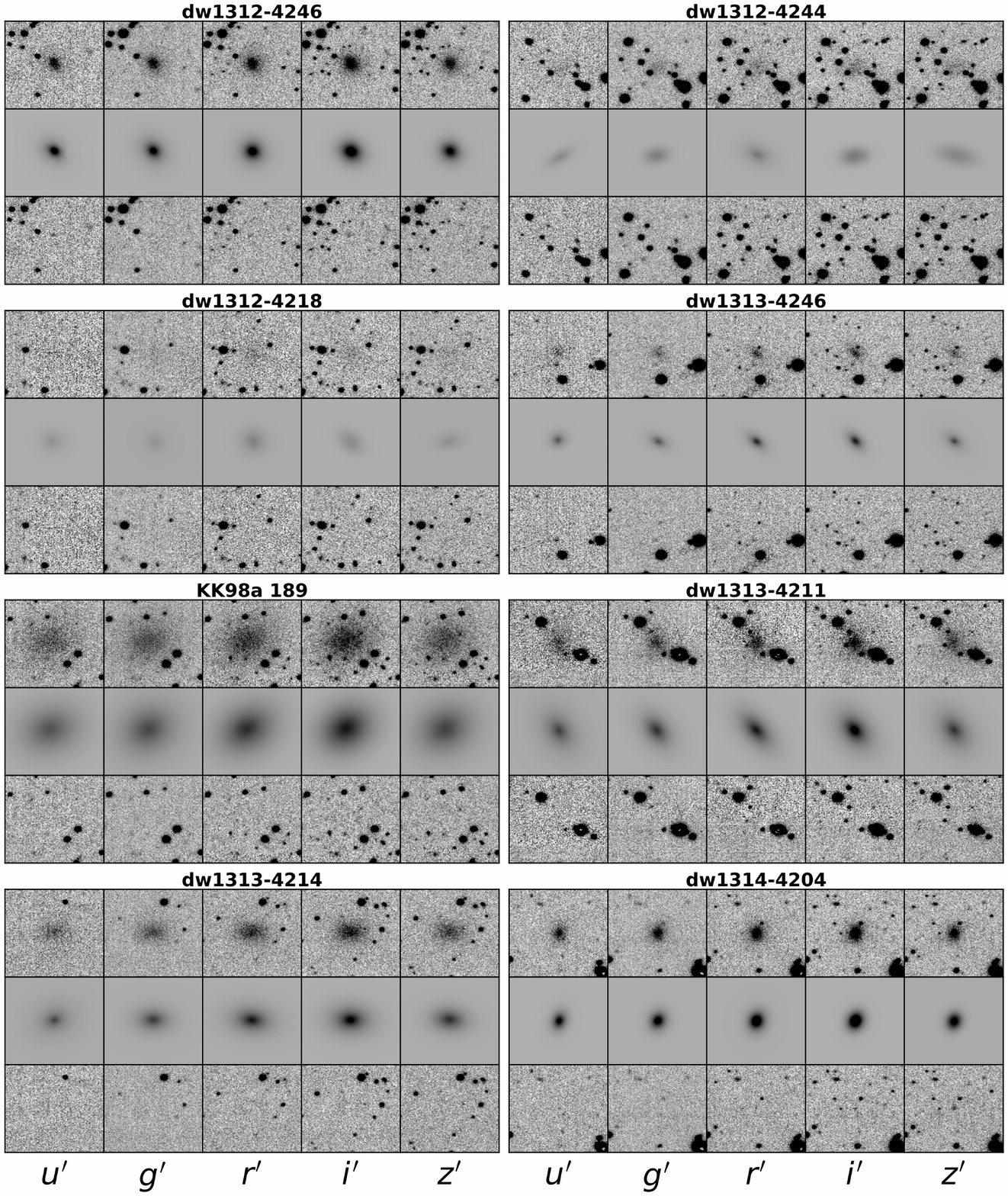}
\caption{Sample of dwarf galaxy model fits.~Blocks are titled by the dwarf identifiers, and show results in the $u'g'r'i'z'$ filters from left to right.~All images measure $\sim\!55\arcsec$ ($\sim1\,{\rm kpc}^2$) on a side. Upper rows in each block show original imaging, with the {\sc galfit} models in the central rows, and model-subtracted residual frames in the row below.}
\label{fig:dw_fits}
\end{figure*} 

We use {\sc galfit} \citep[v3.0.5;][]{pen10} to measure the structural and photometric properties of the dwarfs. We first create $3.65\arcmin\!\times\!3.65\arcmin$ ($4\!\times\!4\,{\rm kpc}^2$) image cutouts in all filters centered on each dwarf. We model the dwarfs using 1D S\'ersic \citep[][]{ser63} profiles defined as $I(r)\!=\!I_{\rm eff}\exp\left\{-b_n\!\left[\!\left(r\!/r_{\rm eff}\!\right)^{1/n}\!-\!1\!\right]\!\right\}$, described by effective radius \reff, the intensity $I_{\rm eff}$ at \reff , and parameters defining the model shape $n$ and $b_n$ that are linked such that half of the total model light is contained within \reff.

For each band we use an iterative surface-brightness ($\mu$) model fitting approach \citep[][]{mun15, ord18a, eig18} by starting with reasonable initial guesses for magnitudes and sizes, and run SE on each cutout with a low detection threshold ({\sc detect\_thresh=0.5}) to construct a segmentation map to mask non-dwarf sources.~We then run {\sc galfit} on the masked image and allow all parameters---including the minor-to-major iso-photal axis ratios ($b/a$) and position angles (PA)---to vary simultaneously. In most cases, models quickly converge to a solution, and in the cases that fail we fix the input magnitude and \reff\ to obtain reasonable estimates on $b/a$ and $n$, and then fix those parameters while freeing the magnitudes and sizes for refinement.~Once an initial model solution has been derived, we then subtract it from the original image, use SE to derive an improved mask, use it to generate a more refined dwarf model, and iterate this process until dwarf properties negligibly vary.

The results of this technique are shown in Figure\,\ref{fig:dw_fits} for a selection of dwarfs, where each block is titled by the dwarf identifiers.~In each block, the original monochromatic imaging is shown in the upper rows with the corresponding models in the middle rows above the residual images.~Close inspection of the residuals shows that the models estimate the $\mu$ profiles very well, with the iterative masking strategy effectively decoupling the dwarf signal even in the presence of foreground star contamination.

We list the full suite of photometric measurements in Table\,\ref{tbl:dwarfphot} alongside absolute $V$-band magnitudes ($M_V$ ), \reff, $n$, and stellar mass estimates (\mstar; see \S\,\ref{sec:mstar}).~In all cases, the recovered photometric centroids for the dwarf candidates agree to $<\!1\arcsec$ across filters, providing consistent anchors upon which the remaining model parameters are based.~Total apparent magnitudes fall in the range $22.31\!\ga\!m_{u'}\!/{\rm mag}\!\ga\!18.82$ and $19.97\!\ga\!m_{z'}\!/{\rm mag}\!\ga\!16.40$ in the bluest and reddest filters, corresponding to foreground extinction-corrected \citep{sch11} $-7.42\!\ga\!M_{V}\!/{\rm mag}\!\ga\!-10.82$, using the conversion of \cite{jes05}.~Overall we find filter-dependent shape parameters in the range $0.15\!\leq\!n\!\leq\!2.73$, and $0.27\!\leq\!b/a\!\leq\!1.00$.~Averaging across filters for each dwarf gives shape parameter ranges of $0.37\!\leq\! n\!\leq\!2.14$, and $0.47\!\leq\!b/a\!\leq\!0.89$ with a sample average $\langle n\rangle\!\approx\!0.94$ and $\langle b/a\rangle\!\approx\!0.72$, consistent with typical structural parameters of other dwarf galaxies in the nearby universe \citep[e.g.][]{eig18}.

While {\sc galfit} is excellent for model fitting, it is deficient in error estimation.~For this, we use the ensemble of all five filters to place uncertainty estimates on the overall sizes and magnitudes.~Table\,\ref{tbl:dwarfphot} lists the mean of each set of filter-dependent model \reff\ along with corresponding 1$\sigma$ error budgets.~To estimate errors on the total dwarf magnitudes we use the variance in the sets of modeled \reff, $n$, and $b/a$ to generate new models by fixing each at their max/minimized values, and leaving the magnitudes unconstrained.~We use the resulting sets of six models to determine the set of (\reff, $n$, $b/a$) that produces the brightest/faintest magnitudes and adopt the difference as the uncertainties listed in Table\,\ref{tbl:dwarfphot}.

\subsection{Dwarf Stellar Masses}
\label{sec:mstar}
We compute \mstar\ by fitting the extinction-corrected total dwarf magnitudes to SEDs predicted by simple stellar population (SSP) synthesis models \citep{bru03}.~We allow SSP ages and metallicities in the ranges $1\!-\!15$\,Gyr and $0.0001\!-\!0.05$\,[Z/H].~We calculate the $\chi^2$ goodness of fit for masses predicted between our observed photometric SEDs, and all SSP models.~We assume an average \mstar, weighted by $\exp(-\chi^2)$, which together with the allowed ranges in age/metallicity help to mitigate the age-metallicity degeneracy \citep{zha17}.~The resulting \mstar\ estimates are listed in the last column of Table\,\ref{tbl:dwarfphot}, and we find they occupy a relatively narrow range of $5.17\la\log$\mstar$/M_\odot\la6.48$, with a mean $\log\langle$\mstar$/M_\odot\rangle\!=\!5.73$.~With sizes of $70\la r_{\rm eff}/{\rm pc}\la300$, these objects are broadly consistent with properties of LG dSphs, in particular those that show the highest stellar surface densities, and have many analogues in the Local Volume \citep[LV;][see also Fig.\ref{fig:dw_sizelum}]{chi09, mcc12, mul15, mul17a}.

\begin{deluxetable*}{lccccccccccc}
\tabletypesize{\scriptsize}
\tablecaption{Photometric properties of the dwarf galaxy candidates\label{tbl:dwarfphot}}
\tablehead{
\colhead{ID}	&	\colhead{$\alpha_{J2000}$}	&	\colhead{$\delta_{J2000}$}	&	\colhead{$u'$}		&	\colhead{$g'$}	&	\colhead{$r'$}	&	\colhead{$i'$}	&	\colhead{$z'$}	&	\colhead{$M_{V}$}	&	\colhead{\reff}	&	\colhead{$n$}	&	\colhead{$\log {\cal M}_\star$}	\\
\colhead{}		&	\colhead{[$hh:mm:ss$]}		&	\colhead{[$dd:mm:ss$]}		&	\colhead{[mag]}		&	\colhead{[mag]}		&	\colhead{[mag]}			&	\colhead{[mag]}		&	\colhead{[mag]}		&	\colhead{[mag]}	&	\colhead{[\arcsec]}	&	\colhead{}	&	\colhead{[$M_\odot$]}	}
\startdata
dw1312-4246 &	13:12:10.18 &	-42:46:48.53 &	20.27$\pm$0.11 &	18.76$\pm$0.12	&	18.81$\pm$0.17	&	17.76$\pm$0.19	&	17.86$\pm$0.13	&	-9.37		&	6.23$\pm$0.70		&	1.32$\pm$0.35	&	$5.89^{+0.34}_{-0.36}$	\\
dw1312-4244 &	13:12:10.93 &	-42:44:43.66 &	21.90$\pm$0.26 &	20.26$\pm$0.19	&	19.88$\pm$0.12	&	19.18$\pm$0.33	&	18.75$\pm$0.47	&	-8.12		&	7.51$\pm$1.32		&	0.68$\pm$0.24	&	$5.35^{+0.35}_{-0.38}$	\\
dw1312-4218 &	13:12:22.48 &	-42:18:41.58 &	21.98$\pm$0.18 &	21.23$\pm$0.24	&	20.16$\pm$0.13	&	19.58$\pm$0.12	&	19.97$\pm$0.15	&	-7.54		&	6.39$\pm$0.66		&	0.68$\pm$0.07	&	$5.17^{+0.34}_{-0.35}$	\\
dw1313-4246 &	13:12:42.87 &	-42:46:50.57 &	21.76$\pm$0.11 &	20.32$\pm$0.12	&	20.11$\pm$0.13	&	19.02$\pm$0.13	&	18.92$\pm$0.11	&	-7.95		&	5.39$\pm$0.65		&	1.32$\pm$0.19	&	$5.43^{+0.35}_{-0.49}$	\\
dw1313-4211 &	13:13:34.28 &	-42:11:08.38 &	19.74$\pm$0.09 &	18.49$\pm$0.08	&	18.29$\pm$0.07	&	17.13$\pm$0.08	&	17.14$\pm$0.09	&	-9.79		&	12.00$\pm$0.96	&	1.07$\pm$0.11	&	$6.15^{+0.35}_{-0.44}$	\\
dw1313-4214 &	13:13:36.40 &	-42:14:08.11 &	20.11$\pm$0.12 &	18.78$\pm$0.16	&	18.43$\pm$0.15	&	17.31$\pm$0.16	&	17.61$\pm$0.17	&	-9.58		&	9.69$\pm$1.19		&	1.06$\pm$0.20	&	$6.03^{+0.35}_{-0.40}$	\\
dw1314-4204 &	13:14:08.17 &	-42:04:08.51 &	20.78$\pm$0.05 &	19.18$\pm$0.05	&	18.97$\pm$0.05	&	18.06$\pm$0.05	&	18.22$\pm$0.04	&	-9.11		&	4.47$\pm$0.30		&	0.91$\pm$0.14	&	$5.78^{+0.35}_{-0.35}$	\\
dw1314-4230 &	13:14:21.93 &	-42:30:41.87 &	20.96$\pm$0.20 &	19.35$\pm$0.12	&	19.05$\pm$0.16	&	18.07$\pm$0.09	&	17.86$\pm$0.11	&	-8.96		&	5.92$\pm$0.72		&	0.89$\pm$0.28	&	$5.89^{+0.35}_{-0.56}$	\\
dw1314-4142 &	13:14:44.82 &	-41:42:28.27 &	21.62$\pm$0.09 &	20.30$\pm$0.07	&	20.55$\pm$0.04	&	19.44$\pm$0.04	&	19.79$\pm$0.08	&	-7.74		&	3.89$\pm$0.25		&	1.01$\pm$0.40	&	$5.23^{+0.35}_{-0.30}$	\\
dw1315-4232 &	13:15:02.98 &	-42:32:17.78 &	20.88$\pm$0.21 &	19.58$\pm$0.23	&	19.02$\pm$0.14	&	18.02$\pm$0.18	&	18.32$\pm$0.21	&	-8.91		&	8.44$\pm$1.78		&	1.50$\pm$0.37	&	$5.76^{+0.35}_{-0.42}$	\\
dw1315-4309 &	13:15:33.97 &	-43:09:27.18 &	22.31$\pm$0.12 &	21.17$\pm$0.18	&	20.63$\pm$0.18	&	19.51$\pm$0.15	&	19.31$\pm$0.17	&	-7.42		&	6.05$\pm$0.76		&	0.46$\pm$0.14	&	$5.27^{+0.36}_{-0.51}$	\\
dw1316-4224 &	13:16:42.27 &	-42:24:05.32 &	19.34$\pm$0.22 &	17.77$\pm$0.15	&	17.94$\pm$0.25	&	16.86$\pm$0.20	&	16.84$\pm$0.14	&	-10.32	&	14.34$\pm$2.01	&	2.14$\pm$0.58	&	$6.30^{+0.34}_{-0.35}$	\\
dw1317-4255 &	13:17:48.49 &	-42:55:40.45 &	21.54$\pm$0.22 &	19.70$\pm$0.21	&	19.89$\pm$0.17	&	18.61$\pm$0.25	&	18.73$\pm$0.30	&	-8.46		&	12.93$\pm$1.11	&	0.37$\pm$0.09	&	$5.55^{+0.35}_{-0.31}$	\\
dw1318-4233 &	13:18:05.59 &	-42:33:37.10 &	20.67$\pm$0.57 &	19.47$\pm$0.42	&	18.84$\pm$0.46	&	18.41$\pm$0.38	&	18.82$\pm$0.34	&	-9.14		&	16.78$\pm$3.95	&	0.37$\pm$0.13	&	$5.61^{+0.32}_{-0.42}$	\\
dw1319-4203 &	13:19:21.26 &	-42:03:38.74 &	20.47$\pm$0.05 &	19.12$\pm$0.06	&	19.05$\pm$0.08	&	18.11$\pm$0.04	&	18.11$\pm$0.07	&	-9.15		&	4.91$\pm$0.40		&	0.52$\pm$0.09	&	$5.82^{+0.34}_{-0.34}$	\\
\hline
KK98a\,189 &	13:12:45.23 &	-41:49:55.23 &	18.82$\pm$0.08 &	17.40$\pm$0.04	&	17.35$\pm$0.04	&	16.42$\pm$0.03	&	16.40$\pm$0.04	&	-10.82	&	14.94$\pm$0.68	&	0.73$\pm$0.05	&	$6.48^{+0.35}_{-0.33}$	\\
\enddata			
\tablecomments{Col.\,(1) lists dwarf identifiers based on the on-sky coordinates listed in cols.\,(2-3). Cols.\,(4-8) show the total apparent magnitudes in each filter, followed by the foreground extinction-corrected absolute $V$-band luminosities , effective radii, average S\'ersic index, and stellar mass estimates in cols.\,(9), (10), (11), and (12), respectively.}
\end{deluxetable*}

\subsection{Size-Luminosity Relation}
\label{sec:sizelum}
\begin{figure*}
\centering
\includegraphics[width=\linewidth]{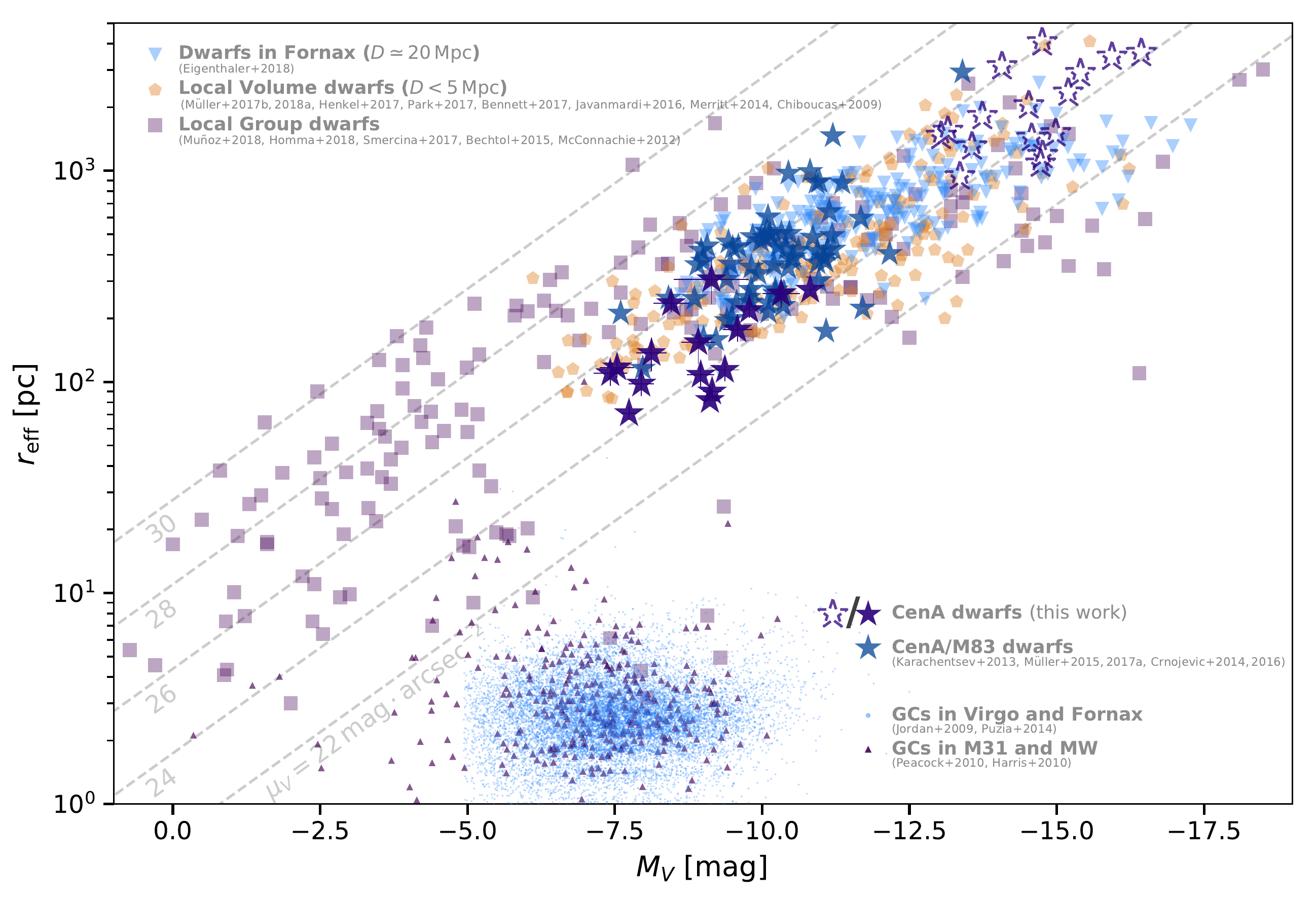}
\caption{The size-luminosity diagram including the present dwarf candidates, LG ultra-faint dwarfs and dSphs, and other dwarf galaxies within the LV and nearby universe, including Fornax cluster dwarfs, and dwarf galaxies associated with more isolated nearby ($\la\!50$\,Mpc; see \S\,\ref{sec:sizelum}) giant galaxies.~We differentiate those already known in the Centaurus\,A/M83 complex for easier comparison to the candidates presented in this work (purple vs.~blue stars).~Where available, we plot $V$-band luminosities, and convert $g'$ and $r'$ magnitudes to $V$ using the relation of \cite{jes05} otherwise, or if not possible plot $g'$ or $r'$-band photometry as available from the literature. Empty dashed stars show the sample from this work shifted to the distance of a background group at a distance of 50\,Mpc (see \S\,\ref{sec:discussion}).
\label{fig:dw_sizelum}}
\end{figure*}

Figure\,\ref{fig:dw_sizelum} compares the modeled \reff\ and total luminosities to other low-mass stellar systems in the \cena-M83 complex \citep[][]{crn14, crn16, mul15, mul17a}, and for additional context to various other dwarfs in the LV, including the M81, M101, M96, and Leo systems \citep[][]{chi09, mer14, jav16, ben17, hen17, par17, mul17b, mul18a}, and the Fornax galaxy cluster core \cite[][]{mun15, eig18}.~We also show low-mass satellites identified in the LG \citep[][]{mcc12, bec15, sme17, hom18, mun18b}.~For consistency with the LG dwarf data, where possible we show absolute $V$-band magnitudes converted from $g'$- and $r'$-band magnitudes \citep{jes05}; however, for some of the dwarfs \cite[][]{chi09, crn16, eig18} either no $g'$ or $r'$ imaging is available, so we defer to those available. 

The size-luminosity relation of the new candidates shows a similar slope as other low-luminosity dwarfs in the nearby universe.~Our selection technique biases us toward higher $\mu_V$ such that the new candidates fall almost parallel to lines of iso-$\mu_V$ (grey dashed lines); a trend that is not replicated by LG systems.~The faintest of the new dwarfs populate a parameter space mostly devoid of LG analogues, but with several representative systems found throughout the LV.~We note that four candidates---namely dw1312-4247, 1314-4204, 1314-4231, and 1319-4203---fall below the main \reff--$M_V$ relation with \reff~$\!\simeq\!100$\,kpc and $M_V\!\simeq\!-9$\,mag.~These objects have no known analogues and we thus consider the possibility of them being associated with background galaxies at a distance of $\sim\!50$\,Mpc. Should this be the case, we indicate the size-luminosity parameter locations for the whole sample by dashed star symbols.~We discuss this notion further in \S\,\ref{sec:discussion}, where we find them more likely to be associated with \cena.

Many of the new candidates coincide with the faintest dwarf galaxies known in the Fornax galaxy cluster at $M_V\!\simeq\!-8$\,mag \citep{eig18}, and overlap with several dwarf galaxies elsewhere in the LV; however, a sharp truncation in the size-luminosity relation is seen beyond the faintest dwarf candidates, with a noticeable gap near $-7.5\la M_V/{\rm mag}\la -5.5$.~We refrain from speculating on the dearth of LG dwarfs in this region, but note that such compact dwarfs beyond the LG can exhibit similar morphologies to background elliptical galaxies in monochromatic imaging, and only reveal themselves via RGB imaging with sufficiently wide SED coverage.~Given this, and the overall difficulty identifying such objects beyond the Local Group, it is likely that this region will continue to be filled out as future imaging campaigns probe ever deeper into various giant galaxy environments.

\begin{figure*}
\centering
\includegraphics[width=\linewidth]{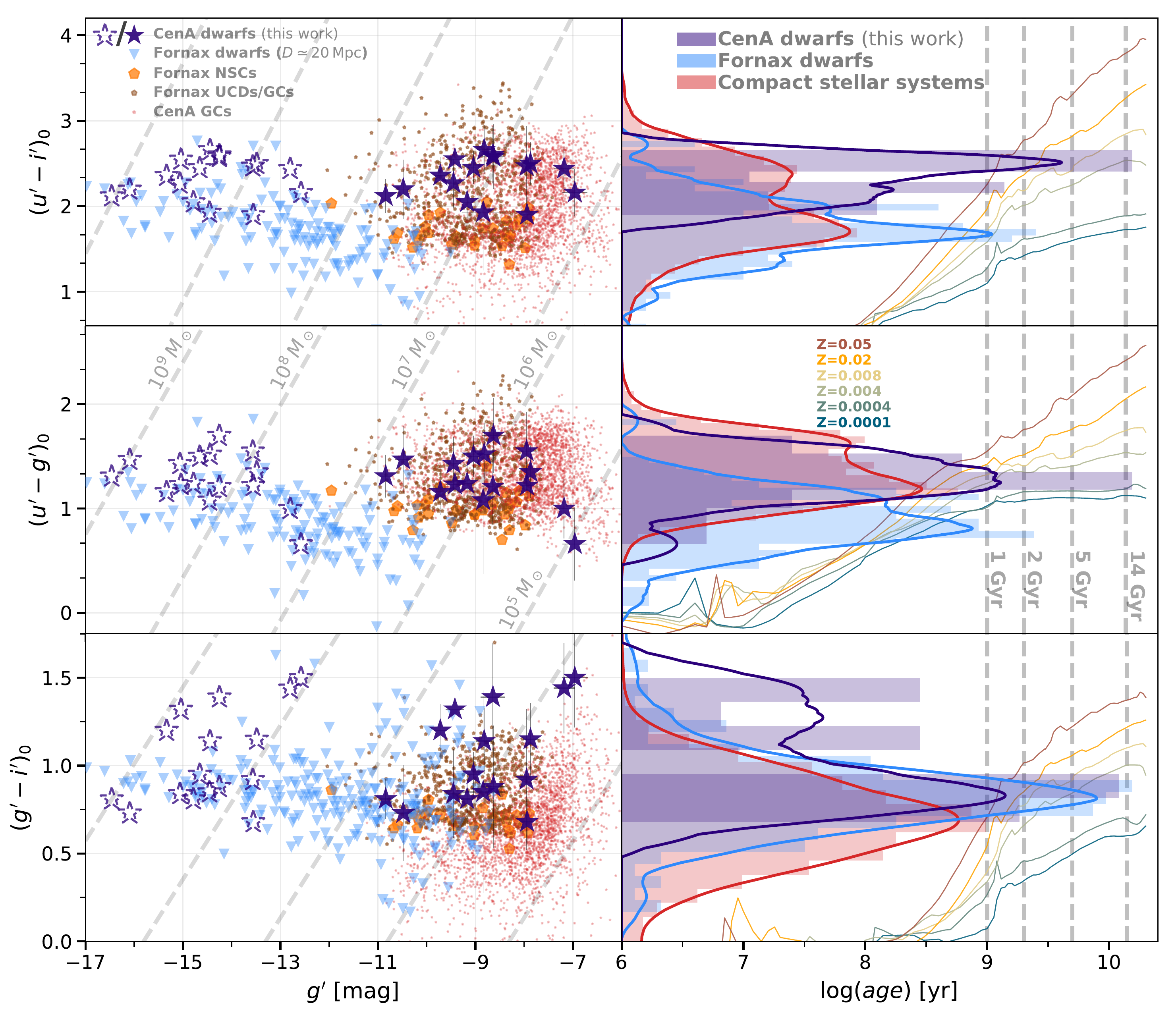}
\caption{Photometric color properties of low-mass stellar systems.~({\it Left panels}):\,$(u'-i')_0$, $(u'-g')_0$, and $(g'-i')_0$ vs.\ $g'$ color-magnitude diagrams are shown from top to bottom, where the current sample is compared to the Fornax cluster sample shown in Fig.\,\ref{fig:dw_sizelum} (blue triangles), where respective photometry is available.~Purple stars show the new \cena\ dwarfs and dashed grey lines show iso-\mstar\ relations using the prescription of \cite{bel03}.~Also shown by orange and brown points are color information for Fornax cluster CSSs including NSCs, UCDs, and GCs, with red points representing \cena\ GCs.~({\it Right panels}): Color distributions are shown aligned along the y-axis for the Fornax (blue), \cena\ (purple), and combined CSS (red) samples and compared to \cite{bru03} SSP models for a range of metallicities shown by the colored tracks.~SSP model tracks are indicated by the thin curves colored according to metallicity, with model ages corresponding to the logarithmic x-axis, and solid grey lines indicating 1, 2, 5, and 14\,Gyr.}
\label{fig:dw_comp}
\end{figure*} 

\subsection{Comparison to Stellar Population Synthesis Predictions}
\label{sec:popsynth}
Figure\,\ref{fig:dw_comp} shows a comparison of the new dwarf candidates to those in the Fornax galaxy cluster with $u'$, $g'$, and $i'$ imaging \citep[][]{mun15,eig18}, with the same symbol/color definitions as in Figure~\ref{fig:dw_sizelum}.~We also plot nuclear star clusters (NSCs) corresponding to the nucleated dwarf galaxies present in the Fornax sample \citep[orange;][]{ord18b}, and compact stellar systems (CSSs) including globular clusters (GCs) and ultra-compact dwarf galaxies (UCDs) also confirmed to be members of the Fornax cluster \citep[brown;][]{wit16}.~Finally, we also show for comparison a large sample of \cena\ GCs \citep[red points;][]{tay17}

The left column shows $(u'-i')_0$, $(u'-g')_0$, and $(g'-i')_0$ vs.\ $M_{g'}$ color-magnitude diagrams (CMDs) for the various samples.~The trend toward bluer colors at fainter luminosities is apparent for the Fornax dwarfs in all three panels, most prominently for the $(u'-i')_0$ CMD, and is typically considered to represent a mass-metallicity relation.~The new \cena\ dwarf candidates are offset from the Fornax dwarf sample toward redder average colors by $\Delta(u'-i')_0\!\approx\!0.6$ and $\Delta(g'-i')_0\!\approx\!0.4$\,mag, and appear more consistent---in terms of stellar population properties---with the nuclei and CSS populations, despite having much more diffuse morphologies.

The right-hand column in Figure~\ref{fig:dw_comp} shows a comparison of the sample color distributions, indicated by the shaded histograms aligned to the y-axes, and corresponding Epanechnikov-kernel density estimates (thicker solid lines).~We find that while there is some overlap with the generally older and metal-poor Fornax dwarf galaxies, the bulk of the \cena\ dwarfs have colors consistent with the secondary red peaks shown by the CSS samples, which shows up most prominently in the $(u'-i')_0$ color. If the redder colors exhibited by the \cena\ dwarfs are due to a metallicity effect, then this might imply early formation within the halos of the giant galaxy progenitors of \cena\ itself, where they could incorporate material rapidly enriched by the giants at early times. Alternatively, higher metallicities could be due to either a prolonged primordial star-formation history enabling self-enrichment, or a non-primordial burst of star formation may have occurred, possibly by previously enriched gas shocked upon infall into \cena's halo. The former scenario might be expected from the density-morphology relation \citep[][]{dre80}, where early star formation failed to be suppressed owing to the relatively low-density environment of the Centaurus\,A group precluding a high number of harassing encounters.~In either case, it would require that these dwarfs must have at one point been embedded in dark matter halos of sufficient mass to retain enriched material expelled by early SNe, which was likely subsequently stripped during infall, thus preventing very recent star formation that would give rise to bluer colors.

\section{Discussion and Summary} \label{sec:discussion}
We have increased the population of likely dwarf galaxy satellites of the Centaurus\,A group by 15, and recovered a single known dwarf \citep{kar98,kar13} in the five optical $u'g'r'i'z'$ bands.~All candidates reside $\sim\!100\!-\!225$\,kpc in \cena's Northwest halo, and their relatively small sizes (\reff~$\!\approx\!70-300$\,pc) and stellar masses (\mstar$\approx\!10^{5-6}\,M_\odot$) are mostly similar to dSphs found in the Local Universe; however, their relatively high $\mu_{\rm eff}\!\approx\!23-26\,{\rm mag}\,{\rm arcsec}^{-2}$ partially populate a region of the low-mass size-luminosity relation not seen in the LG. Rather, these dwarf galaxies appear to be a natural extension of the size-luminosity relation seen in larger galaxy complexes in the nearby universe toward fainter magnitudes not yet well sampled in these complexes.

The relatively high $\mu_V$ may indicate that, rather than being members of the Centaurus\,A group, these candidates may be associated with known giant galaxies in the background. To guard against this, we queried NED for any background sources classified as galaxies falling within 30\arcmin\ of a given dwarf candidate that have measured redshifts and/or distances.~This projected radius corresponds to a physical separation of $\sim200$\,kpc from a giant host located 25\,Mpc away---close enough that dwarf galaxies would be easily spotted by their diffuse natures, while orbiting within a purported host-centric radius where the projected surface number density profile remains high \citep[e.g.][]{ord18a}. Through this exercise, we cannot formally exclude the possibility that at least some, particularly those lying in the Southwestern region of the imaging may be associated with the galaxy group NGC\,5011, located at a distance of $\sim50$\,Mpc. Given this potential host, we calculate the corresponding sizes and luminosities of our candidates should they indeed be members of NGC\,5011.~We plot the results on Figs.\,\ref{fig:dw_sizelum} and \ref{fig:dw_comp} as empty dashed stars, and find that, while the faintest half of our candidates fall within the locus of known dwarf galaxies at the bright/large end of the size-luminosity relation, the brightest show a combination of size and luminosity that is essentially devoid of analogues in the LV.

When we apply the same exercise to the CMDs in Fig.~\ref{fig:dw_comp}, we find that this implies luminosities similar to the brightest dwarf galaxies reported in the Fornax sample, but such luminous dwarf galaxies are simply not found in other LV galaxy complexes that are---like \cena---significantly less massive hosts than Fornax.~Moreover, if a significant number of these candidates are in fact associated with NGC\,5011, this would imply a very shallow slope for the NGC\,5011 faint-end galaxy luminosity function.~While we cannot formally exclude this possibility, given the implication that a group like NGC\,5011 would be hosting a significant population of luminous dwarfs only found so-far in galaxy cluster environments suggests that it is likely that most, if not all, of the present candidates are indeed associated with \cena, which will ultimately require spectroscopic verification.	

Given that the majority of new dwarf candidates are very likely to be associated with \cena, we turn our attention back to Fig.\,\ref{fig:dw_images}, specifically noting the existence of at least two groups of three galaxies (dw1314-4204, 1313-4211, and 1313-4214; and dw1312-4244, 1312-4246, and 1313-4246), each with projected separations of $\la\!20$\,kpc.~While it is impossible to determine their true 3D locations in \cena's halo with the current data, we note that such groups are to be expected from modern cosmological zoom-in simulations \citep[e.g.][]{wet15,bes18}, and thus may represent important examples of dwarf groups in the early stages of interaction while infalling upon \cena's halo.~If so, then these groups, and the present dwarf sample as a whole may represent a valuable opportunity to study these processes in detail in the very nearby universe.

\acknowledgments
We thank the anonymous referee for a fair critique of this work, and the useful feedback that served to improve the original manuscript.

M.A.T.\ is supported by the Gemini Observatory, which is operated by the Association of Universities for Research in Astronomy, Inc., on behalf of the international Gemini partnership of Argentina, Brazil, Canada, Chile, and the United States of America.~This project is supported by FONDECYT Regular Project No.~1161817 and the BASAL Center for Astrophysics and Associated Technologies (PFB-06).

This project used data  obtained with the Dark Energy Camera (DECam), which  was constructed by the Dark Energy Survey  (DES) collaboration.

This research has made use of the NASA Astrophysics Data System Bibliographic Services, the NASA Extragalactic Database, and the SIMBAD database and VizieR catalog access tool, operated at CDS, Strasbourg, France \citep{wen00}. \\

%

\vspace{5mm}
\facilities{CTIO:Blanco/DECam}


\software{{\sc astropy} \citep{ast13},
	  {\sc matplotlib} \citep{hun07},
	  {\sc scamp} \citep{ber02},
	  {\sc swarp} \citep{ber06},  
          {\sc Source Extractor} \citep{ber96},
          {\sc galfit} \citep{pen10},
          }

\end{document}